\RequirePackage{lineno}
\documentclass[preprint, longabstract]{aastex}
\usepackage{natbib}

\begin{document}

\linenumbers
\title{MULTIWAVELENGTH OBSERVATIONS OF THE  AGN 1ES 0414+009 WITH VERITAS, $\textit{Fermi}$-LAT, $\textit{$\textit{Swift}$}$-XRT, AND MDM}
\author{
E.~Aliu\altaffilmark{1},
S.~Archambault\altaffilmark{2},
T.~Arlen\altaffilmark{3},
T.~Aune\altaffilmark{4},
M.~Beilicke\altaffilmark{5},
W.~Benbow\altaffilmark{6},
M.~B{\"o}ttcher\altaffilmark{7},
A.~Bouvier\altaffilmark{4},
V.~Bugaev\altaffilmark{5},
A.~Cannon\altaffilmark{8},
A.~Cesarini\altaffilmark{9},
L.~Ciupik\altaffilmark{10},
E.~Collins-Hughes\altaffilmark{8},
M.~P.~Connolly\altaffilmark{9},
W.~Cui\altaffilmark{11},
R.~Dickherber\altaffilmark{5},
J.~Dumm\altaffilmark{12},
M.~Errando\altaffilmark{1},
A.~Falcone\altaffilmark{13},
S.~Federici\altaffilmark{14,15},
Q.~Feng\altaffilmark{11},
J.~P.~Finley\altaffilmark{11},
G.~Finnegan\altaffilmark{16},
L.~Fortson\altaffilmark{12},
A.~Furniss\altaffilmark{4},
N.~Galante\altaffilmark{6},
D.~Gall\altaffilmark{17},
S.~Godambe\altaffilmark{16},
S.~Griffin\altaffilmark{2},
J.~Grube\altaffilmark{10},
G.~Gyuk\altaffilmark{10},
D.~Hanna\altaffilmark{2},
H.~Huan\altaffilmark{18},
G.~Hughes\altaffilmark{14},
C.~M.~Hui\altaffilmark{16},
A.~Imran\altaffilmark{19},
O.~Jamil\altaffilmark{7},
P.~Kaaret\altaffilmark{17},
N.~Karlsson\altaffilmark{12},
M.~Kertzman\altaffilmark{20},
J.~Kerr\altaffilmark{7},
Y.~Khassen\altaffilmark{8},
D.~Kieda\altaffilmark{16},
H.~Krawczynski\altaffilmark{5},
F.~Krennrich\altaffilmark{19},
M.~J.~Lang\altaffilmark{9},
K.~Lee\altaffilmark{5},
A.~S~Madhavan\altaffilmark{19},
P.~Majumdar\altaffilmark{3},
S.~McArthur\altaffilmark{5},
A.~McCann\altaffilmark{2},
P.~Moriarty\altaffilmark{21},
R.~Mukherjee\altaffilmark{1},
T.~Nelson\altaffilmark{12},
A.~O'Faol\'{a}in de Bhr\'{o}ithe\altaffilmark{8},
R.~A.~Ong\altaffilmark{3},
M.~Orr\altaffilmark{19},
A.~N.~Otte\altaffilmark{22},
N.~Park\altaffilmark{18},
J.~S.~Perkins\altaffilmark{23,24},
A.~Pichel\altaffilmark{25},
M.~Pohl\altaffilmark{26,14},
J.~Quinn\altaffilmark{8},
K.~Ragan\altaffilmark{2},
P.~T.~Reynolds\altaffilmark{27},
E.~Roache\altaffilmark{6},
J.~Ruppel\altaffilmark{26,14},
D.~B.~Saxon\altaffilmark{28},
M.~Schroedter\altaffilmark{6},
G.~H.~Sembroski\altaffilmark{11},
G.~D.~\c{S}ent\"{u}rk\altaffilmark{29},
A.~W.~Smith\altaffilmark{16,*},
D.~Staszak\altaffilmark{2},
M.~Stroh\altaffilmark{17},
I.~Telezhinsky\altaffilmark{26,14},
G.~Te\v{s}i\'{c}\altaffilmark{2},
M.~Theiling\altaffilmark{11},
S.~Thibadeau\altaffilmark{5},
K.~Tsurusaki\altaffilmark{17},
A.~Varlotta\altaffilmark{11},
V.~V.~Vassiliev\altaffilmark{3},
M.~Vivier\altaffilmark{28},
S.~P.~Wakely\altaffilmark{18},
J.~E.~Ward\altaffilmark{5},
A.~Weinstein\altaffilmark{19},
R.~Welsing\altaffilmark{14},
D.~A.~Williams\altaffilmark{4},
B.~Zitzer\altaffilmark{30}
}
\altaffiltext{*}{Corresponding Author: aw.smith@utah.edu}
\altaffiltext{1}{Department of Physics and Astronomy, Barnard College, Columbia University, NY 10027, USA}
\altaffiltext{2}{Physics Department, McGill University, Montreal, QC H3A 2T8, Canada}
\altaffiltext{3}{Department of Physics and Astronomy, University of California, Los Angeles, CA 90095, USA}
\altaffiltext{4}{Santa Cruz Institute for Particle Physics and Department of Physics, University of California, Santa Cruz, CA 95064, USA}
\altaffiltext{5}{Department of Physics, Washington University, St. Louis, MO 63130, USA}
\altaffiltext{6}{Fred Lawrence Whipple Observatory, Harvard-Smithsonian Center for Astrophysics, Amado, AZ 85645, USA}
\altaffiltext{7}{Astrophysical Institute, Department of Physics and Astronomy, Ohio University, Athens, OH 45701}
\altaffiltext{8}{School of Physics, University College Dublin, Belfield, Dublin 4, Ireland}
\altaffiltext{9}{School of Physics, National University of Ireland Galway, University Road, Galway, Ireland}
\altaffiltext{10}{Astronomy Department, Adler Planetarium and Astronomy Museum, Chicago, IL 60605, USA}
\altaffiltext{11}{Department of Physics, Purdue University, West Lafayette, IN 47907, USA }
\altaffiltext{12}{School of Physics and Astronomy, University of Minnesota, Minneapolis, MN 55455, USA}
\altaffiltext{13}{Department of Astronomy and Astrophysics, 525 Davey Lab, Pennsylvania State University, University Park, PA 16802, USA}
\altaffiltext{14}{DESY, Platanenallee 6, 15738 Zeuthen, Germany}
\altaffiltext{15}{Institut f\"{u}r Physik und Astronomie, Universit\"{a}t Potsdam, 14476 Potsdam-Golm,Germany; DESY, Platanenallee 6, 15738 Zeuthen, Germany}
\altaffiltext{16}{Department of Physics and Astronomy, University of Utah, Salt Lake City, UT 84112, USA}
\altaffiltext{17}{Department of Physics and Astronomy, University of Iowa, Van Allen Hall, Iowa City, IA 52242, USA}
\altaffiltext{18}{Enrico Fermi Institute, University of Chicago, Chicago, IL 60637, USA}
\altaffiltext{19}{Department of Physics and Astronomy, Iowa State University, Ames, IA 50011, USA}
\altaffiltext{20}{Department of Physics and Astronomy, DePauw University, Greencastle, IN 46135-0037, USA}
\altaffiltext{21}{Department of Life and Physical Sciences, Galway-Mayo Institute of Technology, Dublin Road, Galway, Ireland}
\altaffiltext{22}{School of Physics \& Center for Relativistic Astrophysics, Georgia Institute of Technology, 837 State Street NW, Atlanta, GA 30332-0430}
\altaffiltext{23}{CRESST and Astroparticle Physics Laboratory NASA/GSFC, Greenbelt, MD 20771, USA.}
\altaffiltext{24}{University of Maryland, Baltimore County, 1000 Hilltop Circle, Baltimore, MD 21250, USA.}
\altaffiltext{25}{Instituto de Astronomia y Fisica del Espacio, Casilla de Correo 67 - Sucursal 28, (C1428ZAA) Ciudad Aut—noma de Buenos Aires, Argentina}
\altaffiltext{26}{Institut f\"ur Physik und Astronomie, Universit\"at Potsdam, 14476 Potsdam-Golm,Germany}
\altaffiltext{27}{Department of Applied Physics and Instrumentation, Cork Institute of Technology, Bishopstown, Cork, Ireland}
\altaffiltext{28}{Department of Physics and Astronomy and the Bartol Research Institute, University of Delaware, Newark, DE 19716, USA}
\altaffiltext{29}{Physics Department, Columbia University, New York, NY 10027, USA}
\altaffiltext{30}{Argonne National Laboratory, 9700 S. Cass Avenue, Argonne, IL 60439, USA}

\begin{abstract}
We present observations of the BL Lac object 1ES 0414+009 in the $>$200 GeV gamma-ray band by the VERITAS array of Cherenkov telescopes. 1ES 0414+009 was observed by VERITAS between January 2008 and February 2011, resulting in 56.2 hours of good quality pointed observations. These observations resulted in a detection of 822 events from the source corresponding to a statistical significance of 6.4 standard deviations (6.4 $\sigma$) above the background. The source flux, showing no evidence for variability, is measured as  (5.2$\pm1.1_{stat} \pm 2.6_{sys}) \times$ 10$^{-12}$ photons cm$^{-2}$ s$^{-1}$ above 200 GeV , equivalent to approximately 2$\%$ of the Crab Nebula flux above this energy. The differential photon spectrum from 230 GeV to 850 GeV is well fit by a power law with an photon index  of $\Gamma$= 3.4 $\pm$0.5$_{stat}\pm$0.3$_{sys}$ and a flux normalization of (1.6$\pm$0.3$_{stat}\pm0.8_{sys}$) $\times$ 10$^{-11}$ photons cm$^{-2}$ s$^{-1}$ at 300 GeV. We also present multiwavelength results taken in the optical (MDM), X-ray ($\textit{Swift}$-XRT), and GeV ($\textit{Fermi}$-LAT) bands and use these results to construct a broadband spectral energy distribution (SED). Modeling of this SED indicates that homogenous one-zone leptonic scenarios are not adequate to describe emission from the system, with a lepto-hadronic model providing a better fit to the data.
\end{abstract}

\section{Introduction}
Among the more substantial accomplishments of the latest generation of imaging atmospheric Cherenkov telescopes (IACTs) is the reliable detection of increasingly distant active galactic nuclei (AGN). While the previous generation of ground based gamma-ray experiments such as the Whipple 10 meter telescope and HEGRA detected comparatively close AGN, the furthest being H 1426+428  at a redshift of 0.129 \citep{Horan}, observations with VERITAS, HESS, and MAGIC have shown that it is possible to detect sources of very high energy (VHE) gamma rays ($>$100 GeV) as far out as redshifts of z= 0.54 \citep{3c279}.

The AGN 1ES 0414+009 is among the furthest detected TeV BL Lacertae objects, having a well measured redshift of z=0.287 \citep{0414redshift}. 1ES 0414+009 was first discovered in an X-ray survey performed by the HEAO 1 A-1 instrument \citep{Ulmer} and was originally associated with a cluster of galaxies. Further radio, X-ray and optical observations led to the association of the X-ray source with an X-ray bright BL Lacertae object \citep{Ulmer1983}.~Additional X-ray observations \citep{Wolter,Beckmann} show 1ES 0414+009 to have typical spectral characteristics of ``high frequency peaked BL Lacertae'' (HBL) objects, as well as being comparatively bright in the 2-10 keV band with a luminosity comparable to Markarian 421 and PKS 2155-304, allowing the possibility that 1ES 0414+009 may be a high-redshift analog of these well studied TeV blazars. 

1ES 0414+009 also appears in the $\textit{Fermi}$-LAT 2-year source catalog, associated with the source 2FGL J0416.8+0105\footnote[1]. Its inclusion in the 2FGL catalog is based on its detection in the 1-100 GeV band at a significance of 6.8 $\sigma$; the photon spectrum of the excess is well fit by a power law with a photon index of 1.96$\pm 0.16_{stat}$ and an integral flux in the 1-100 GeV band of (6.9$\pm1.4_{stat}) \times$10$^{-10}$ photons cm$^{-2}$ s$^{-1}$.

1ES 0414+009 was among the 33 objects deemed to be very good candidates for TeV detections in the list of \citet{Costamantelist} and, as such, preliminary observations were taken by several TeV observatories with only upper limits resulting \citep{Whipple0414UL,HEGRA0414UL,MAGIC0414UL}.  The HESS array of IACT telescopes detected the source in observations taken from 2005 to 2009 \citep{HESSATEL}, with 224 excess events above 200 GeV in 73.7 hours of observation, corresponding to a statistically significant excess of 7.8$\sigma$ \citep{TEXAS}. The HESS excess is well fit by a power law with a photon index of 3.5$\pm0.3_{stat}\pm0.2_{sys}$ and an integral flux above 200 GeV of (1.9$\pm0.2_{stat}\pm0.4_{sys}) \times 10^{-12}$ photons cm$^{-2}$ s$^{-1}$ or $\sim$0.6$\%$ of the Crab Nebula flux in the same range.

\section{VERITAS Observations}

The VERITAS array \citep{ONGVERITAS} of IACTs located in southern Arizona (1.3 km m.a.s.l., 31$^{\circ}$40'30''N, 110$^{\circ}$57'07'' W) began 4-telescope array observations in September 2007 and is the most sensitive IACT for observations above 200 GeV currently in operation. The array is composed of four 12m diameter telescopes, each with a Davies-Cotton tessellated mirror structure of 345 12m focal length hexagonal mirror facets (total mirror area of 110 m$^{2}$). Each telescope focuses Cherenkov light from particle showers onto its 499-pixel PMT  (photomultiplier tube) camera. Each pixel has a field of view of 0.15$^{\circ}$, resulting in a camera field of view of 3.5$^{\circ}$. VERITAS has the capability to detect and measure gamma rays in the 100 GeV to 30 TeV energy regime with an energy resolution of 15-20$\%$ and an angular resolution of $<$0.1$^{\circ}$ on an event by event basis. 

Between June and September 2009, VERITAS underwent a significant reconfiguration in which one of the four telescopes was moved to a new location making the geometric configuration of the array more symmetric and increasing the overall baseline \citep{Perkinsrelocate}.~This reconfiguration, as well as a marked improvement in the process for VERITAS mirror alignment \citep{McCannMirror}, yielded a significant improvement in the VERITAS sensitivity at the level of 30$\%$. In its current configuration, VERITAS can detect a 1$\%$ Crab Nebula flux  at a 5$\sigma$ significance in under 30 hours, improved from the $\sim$45 hours required with the previous array configuration and alignment. 

The VERITAS observations of 1ES 0414+009 were made between January 2008 and February 2011 and therefore comprise data taken in both array configurations. After quality selection cuts which remove data taken during poor weather and hardware conditions,  approximately 25$\%$ (12.1 hours) of the data for this analysis was taken with the original configuration and 75$\%$ (44.1 hours) were taken with the new configuration. The observations were made in ``wobble'' mode in which the source is offset from the center of the field-of-view of the cameras to maximize efficiency in obtaining both source and background measurements \citep{Fomin}. All data were processed and analyzed with the VEritas Gamma-ray Analysis Suite (VEGAS, \citet{VEGAS}) with all results being successfully cross-checked by a second independent analysis package \citep{ED}.  The event selection criteria (cuts) used are based on the image morphology (\textit{Mean Scaled Width} and \textit{Length}), and the angular distance between the reconstructed position of the source in the camera plane and the $\textit{a priori}$ known source location \citep{CUTS} and were optimized on a simulated source with a soft energy spectrum ($\Gamma$=4) and a 7$\%$ Crab Nebula flux (at 200 GeV).

From the entire observational sample, a total of 822 events was detected in excess of the estimated background, corresponding to a statistical significance of 6.4$\sigma$ \citep{LIMA}.  A two dimensional Gaussian fit to the VERITAS excess (VER J0416+011) is consistent with a point source located at  04$^{h}$16$^{m}$54$^{s}\pm5^{s}_{stat}\pm6^{s}_{sys}$, +1$^{\circ}$06$^{'}$35$^{''}\pm1'47''_{stat}\pm1'30''_{sys}$ (J2000), and with the optical AGN position \citep{0414OpticalPosition}. VER J0416+011 is shown in Figure 1 along with the $\textit{Fermi}$-LAT and optical positions. 

   The differential photon spectrum (see Figure 2) obtained from the VERITAS observations is well fit by a power law of the form (1.6$\pm$0.3$_{stat}\pm0.8_{sys}$) $\times$ 10$^{-11}$ $\times (\frac{E}{0.3 TeV})^{-3.4\pm0.5_{stat}\pm0.3_{sys}}$cm$^{-2}$s$^{-1}$TeV$^{-1}$ with a reduced chi-squared value ($\chi^{2}$/d.o.f) of 0.57 for 2 degrees of freedom. The VERITAS spectral parameters are consistent with those measured by HESS. 
    
      Due to its relatively low transit path on the sky for VERITAS, the data on 1ES 0414+009 were taken over a small range of zenith angles from 30$^{\circ}$ to 40$^{\circ}$, resulting in an energy threshold \footnote[2]{The energy threshold is defined as the energy corresponding to the maximum of the product function of the observed spectrum with the collection area of the instrument.} of 200 GeV. The integral flux, obtained from 1ES 0414+009 during VERITAS observations (derived from the spectral fit shown in Figure 2) was (5.2$\pm$1.1$_{stat}\pm$2.6$_{sys}$) $\times$10$^{-12}$ photons cm$^{-2}$, or  2$\%$ of the Crab Nebula flux above 200 GeV. The construction of a lightcurve, binned on the timescale of VERITAS ``dark runs" (intervals between full moons when skies are dark) shows no evidence for variability with a straight line fit yielding a reduced $\chi^{2}$ of 0.55 for a chance probability of 88$\%$ (see Figure 3). The HESS observations, which indicated a 0.6$\%$ Crab Nebula flux, were made between October 2005 and September 2009, overlapping only slightly with the VERITAS observations. While the VERITAS integral flux is larger than the HESS measurement, when statistical and systematic errors are taken into account (see \citet{TEXAS}), the two flux measurements are consistent with each other. However, given the current data we cannot rule out some long term variability in flux.

\section{Multiwavelength Observations}
\subsection{$\textit{Fermi}$-LAT}
$\textit{Fermi}$-LAT  \citep{Atwood} analysis was performed on all available photons from 0.3 to 300 GeV accrued from the start of full Fermi-LAT science operations (August 4, 2008) until July 14, 2011 (MJD  54682-55756), overlapping to a large degree with the VERITAS observations (January 2008 - February 2011). The data were analyzed using the ScienceTools v9r23p1 package available from the Fermi Science Support Center (FSSC)\footnote[3]{http://fermi.gsfc.nasa.gov/ssc/}. Standard data quality cuts were applied as recommended by the FSSC, with only ``diffuse" class photons being used for this analysis (events with a high probability of being correctly identified as photons). An un-binned analysis was performed on all suitable photons coming from a region of radius 15$^{\circ}$ centered around 1ES 0414 +009 with all 2FGL sources (sources identified within the $\textit{Fermi}$-LAT 2 year catalog) lying within 15$^{\circ}$ of 1ES 0414+009 being modeled as well. In addition, both Galactic and extragalactic contributions to the diffuse background were modeled out using the P6$\_$V11$\_$Diffuse response functions. The Fermi-LAT systematic uncertainties are conservatively estimated at 10$\%$ for the energy range under evaluation in this work. \footnote[4]{http://fermi.gsfc.nasa.gov/ssc/data/analysis/LAT$\_$caveats.html/}

The source is detected in the 0.3-300 GeV energy range with a test-statistic (TS) of 97 corresponding to a detection significance of approximately 9.8$\sigma$. A lightcurve binned on a monthly timescale shows no evidence for variability in the $\textit{Fermi}$-LAT energy range with a straight line fit yielding a reduced $\chi^{2}$ of 0.81 for a chance probability of 76$\%$(see Figure 3). The source spectrum is well fit by a power law of the form (6.3$\pm$ 1.5$_{stat}$) $\times$10$^{-13}$$(\frac{E}{1000 MeV})^{-1.9\pm{0.1_{stat}}}$ photons cm$^{-2}$s$^{-1}$MeV$^{-1}$. For modeling purposes, spectral data points were obtained by dividing the data into five equally spaced bins in log(E) and requiring that each bin have a TS value of at least 4. These spectral points along with the associated fit to the $\textit{Fermi}$-LAT data are shown in Figure 2. As can be seen, the spectra from the two instruments ($\textit{Fermi}$-LAT and VERITAS) connect to each other quite well, with the attenuating effect of the extragalactic background light on the observed TeV flux being readily apparent. It should be noted that the two spectra were taken over different (although overlapping), extended time periods and cannot be considered to constitute a simultaneous measurement of the GeV-TeV spectrum.

Using the ScienceTools function $\textit{gtprobsrc}$, the excess seen in the $\textit{Fermi}$-LAT data is consistent with a point source located at RA = 04$^{h}$16$^{m}$51$^{s}\pm8^{s}_{stat}$, DEC = +1$^{\circ}$04$^{'}$36$^{''}\pm2'2''_{stat}$ (J2000), consistent within errors with both the AGN position from \citet{0414OpticalPosition} and the VERITAS source location (see Figure 1). It  should be noted that all $\textit{Fermi}$-LAT analysis results presented here are consistent with those derived for the source 2FGL J0416.8+0105 in the 2 year $\textit{Fermi}$-LAT point source catalog.

\subsection{\textit{$\textit{Swift}$}-XRT}

Between 2006 and 2009, a total of 6.24 ks of $\textit{Swift}$-XRT  \citep{Burrows} observations were taken on 1ES 0414+009 in ``photon counting" (PC) mode. The reduced data products used for this study were downloaded from the UK $\textit{Swift}$ Science Center\footnote[5]{http://www.swift.ac.uk/} \citep{SwiftUK}. Calibrated source and background files, along with the appropriate response files, were used as inputs to the spectral fitting package XSPEC v12.6 \citep{XSPEC}. The spectrum between 0.3 and 10 keV was fit while allowing the neutral hydrogen (HI) column density to vary freely.  The resulting HI column density  was fit as (1.7 $\pm$ 0.16) $\times$ 10$^{21}$ cm$^{-2}$,  higher than the value of 0.85 $\times$ 10$^{22}$ cm$^{-2}$  quoted in \citet{Kaberla}.  The spectrum is well fit (C-statistics were applied with less than 5$\%$ of model realizations providing a better fit) by a simple, absorbed power law  (accounting for the redshift of the target) using the Wisconsin photoelectric cross sections \citep{WABS} or the ``zwabs'' model in XSPEC. The best fit for this model is provided by a normalization of (6.1 $\pm 0.3) \times10^{-3}$ ph cm$^{-2}s^{-1}$ at 1 keV and a photon index of -2.4$\pm0.1$. All of these results are consistent with an analysis of the individual pointings analyzed separately, presented in Abramowski (2012). The deabsorbed source flux is measured as (20.3 $\pm 0.1) \times10^{-12}$erg cm$^{-2}$s$^{-1}$ with no evidence for strong variability in the PC mode data presented here.  

It should be noted, however, that $\sim$2200 seconds of Swift-XRT observations taken in Ówindowed timingÓ (WT) mode on MJD 55231 and 55241 in 2010 indicate a large increase in the 0.3-10 keV flux from 1ES 0414+009. This data is fit by a simple, absorbed power-law with a derived photon index entirely consistent with the index from the PC mode observations, while showing a 0.3-10 keV flux of 58.8 ($\pm 0.2)  \times10^{-12}$erg cm$^{-2}$s$^{-1}$, a factor of 2-3 higher than previous observations. Since the WT mode observations suggest a relatively short flaring episode, while the rest of the multi-wavelength data were taken over a much larger timescale, the WT mode observations are not included in the modeling effort in this work.  However, we report these flare data as an indication of overall variable behavior in the system.

\subsection{MDM}
1ES 0414+009 was observed at the 1.3m McGraw-Hill Observatory of the Michigan-Dartmouth-MIT (MDM) Observatory\footnote[6]{http://mdm.kpno.noao.edu/} on the southwest ridge of Kitt Peak, during the nights of January 26 - 30, 2011. Standard V, R, and I filters were used. The data were bias-subtracted and flat-field corrected using
standard routines in IRAF, and comparative photometry was done based on comparison star magnitudes from \citep{Fiorucci}. For the construction of VRI SEDs, the magnitudes
were corrected with the extinction coefficients A$_{V }$ = 0.39, A$_{R}$ = 0.32, and A$_{I}$ = 0.23, as provided by the NASA/IPAC Extragalactic Database\footnote[5]{http://ned.ipac.caltech.edu/ } (NED). In addition to the MDM observations, we include archival U,B,V,R, and I filter data from \citet{AIT} and \citet{Ulmer1983}; along with infrared J, H, K values taken from \cite{2MASS}.

\section{Multiwavelength SED and Modeling}

The broadband SED constructed from the observations of 1ES 0414+009 is shown in Figure 4. Modeling of this SED proceeded by applying three distinct models of emission (two purely leptonic and one lepto-hadronic hybrid). The resulting predictions of the gamma-ray emission from these models were then modified to account for the attenuating effect of the extragalactic background light (EBL) using  the \citet{Finke} EBL model. 

The purely leptonic models employed to fit the SED are based on the premise that the high energy (GeV-TeV) emission is produced by inverse-Compton scattering of low energy ambient photons by a population of relativistic jet electrons (whose primary synchrotron emission makes up the lower energy X-ray peak). The variations between the models are based upon the nature of the photon field which is up-scattered: in synchrotron self-Compton (SSC) models, the target photon field is made of the original primary synchrotron photons (from the lower energy peak); while in external Compton (EC) models the emission includes the SSC contribution as well as inverse-Compton emission from a target photon field external to the jet.

The SSC model employed here is that of \citet{Bottchermod} described also in detail in \citet{WCom}. In this model, the emission originates from a spherical distribution of relativistic electrons (of radius R) traveling along the jet axis with a Lorentz factor $\Gamma$, corresponding to a jet speed of $\beta_{\Gamma}$c. The jet viewing angle $\Theta_{obs}$ results in a Doppler boost factor given by D = ($\Gamma[1-\beta_{\Gamma}cos\Theta_{obs}])^{-1}$.  In order to reduce the number of free parameters, we choose $\Theta_{0bs}$ = 1/$\Gamma$, leading to $\Gamma = D$.

In models such as these, particles are injected with some distribution in energy and are subsequently cooled (i.e. lose energy by radiative mechanisms) or escape the accelerating region and no longer contribute to emission. The models under consideration here result in an equilibrium between these factors (i.e. acceleration and cooling/escape). Here we parameterize the front end of the model by injecting electrons into the spherical region with a power-law distribution given by Q($\gamma$) = Q$_{0}\gamma^{-q_{e}}$ bounded by the cutoff parameters $\gamma_{emin},\gamma_{emax}$. The escape time in the model is parameterized by $\eta$ which corresponds to an escape time t$_{esc}$ of the particles of t$_{esc}$ = $\eta$R/c. The resulting equilibrium particle distribution in the emission volume corresponds to a kinetic power L$_{e}$ in relativistic electrons. The synchrotron emission is determined by a tangled magnetic field B, corresponding to a power L$_{B}$ in Poynting flux. Furthermore, the equipartition parameter $\epsilon_{eB}$ = L$_{B}$/L$_{e}$ is evaluated during the modeling process to give an evaluation of how reasonable the model is physically; it is reasonable to expect that most models that fit HBL emission well would generate values for $\epsilon_{eB}$ within the range of $\sim$0.1-1.0.

The EC model used in this analysis incorporates all the above SSC model attributes while adding an additional emission component due to inverse-Compton scattering off an external photon source (corresponding to a standing perturbation of the jet flow). Hence its modeling parameters are identical to the SSC parameters along with the addition of the blackbody temperature and energy density of the external photon field (T$_{EC}$, $\textit{u}_{EC}$).

In contrast to the two leptonic models above, the lepto-hadronic model used for this analysis assumes that the high energy emission is dominated by proton synchrotron and pion decay emission processes. An additional model component is generated from the electromagnetic cascades from $\gamma-\gamma$ absorption between multi-TeV photons (from charged-pion decay products) and low-energy synchrotron photons from ultrarelativistic leptons. Similarly to the SSC and EC models, the protons are injected with a power-law spectrum (parameterized by $\textit{n}$($\gamma$) = $\gamma^{-q_{p}}$ bracketed by the lower and higher energy cutoffs $\gamma_{pmin},\gamma_{pmax}$) corresponding to a kinetic power L$_{p}$ in relativistic protons, which we compare with the power L$_{e}$ in relativistic primary leptons and the Poynting flux power L$_{B}$. All parameter values described here, as well as resulting equipartition values, are listed in Table 1.

As can be seen in Figure 4, while both leptonic models fit the TeV spectrum reasonably well, neither provides a satisfactory fit to the observed broad GeV peak (in contrast to the relatively sharp synchrotron peak).  The extremely wide separation between the synchrotron and gamma-ray peak implies a very large $\gamma$ of the electrons (see Table 1), which then implies a very low B-field in order to achieve the observed synchrotron peak. A large Doppler factor (D = 40) can then help reduce the overall power requirements in the electrons to maintain values close to reasonable equipartition conditions.  

We find that the lepto-hadronic model provides a more satisfactory fit to the observed SED. The primary proton synchrotron emission matches the GeV - TeV emission very well, while the lower end of the GeV spectrum is reproduced by emission from electromagnetic cascades. It is clear however, that this model still over/underpredicts some of the $\textit{Fermi}$-LAT GeV points and is far from equipartition. It is possible that additional components may be required in this model to more accurately describe the system. Additionally, the model predicts a relatively strong flux in the very hard X-ray/soft gamma-ray regime; an energy range which is completely unexamined by the observations described in this work.

\section{Summary}

The VERITAS array of telescopes has detected the distant blazar 1ES0414+009 in 56.2 hours of observations, resulting in a detection at a statistical significance of 6.4$\sigma$. The differential photon spectrum from the source is well fit by a power law with normalization and photon index parameters given by (1.6$\pm$0.3$_{stat}\pm0.8_{sys}$) $\times$ 10$^{-11}$ $\times \frac{E}{0.3 TeV}^{-3.4\pm0.5_{stat}\pm0.3_{sys}}$cm$^{-2}$s$^{-1}$TeV$^{-1}$. The integral flux from the source is measured as approximately 2$\%$ of the Crab Nebula flux above 200 GeV. While this is a larger integral flux than the HESS measurement (0.6$\%$ Crab above 200 GeV), when statistical and systematic errors are taken into account (see \citet{TEXAS}), the two flux measurements are consistent with each other. 

We have also presented multiwavelength data on 1 ES 0414+009 from $\textit{Fermi}$-LAT, $\textit{Swift}-XRT$, and the MDM observatory. We combine these data with archival optical data to construct a broadband SED which shows a very narrow synchrotron peak in contrast to the relatively broad GeV-TeV gamma-ray peak. The SED is not well fit by homogenous leptonic one-zone models, however a hybrid lepto-hadronic model provides a reasonable fit to the observed data in a strongly magnetically dominated jet. 

Although the lepto-hadronic model provides a better fit to the observed SED, it still does not accurately predict the apparently complicated form of the GeV-TeV emission. It should be noted, however,  that both the GeV and TeV spectral data points have relatively large error bars and do not take systematic errors into account. Therefore, it is difficult to accurately constrain or improve any SED models until much more precise measurements of the GeV-TeV regime emission are made. 

The study of sources at such a relatively high redshift can shed light on more fundamental science topics such as the characteristics of the extragalactic background light (EBL) and intergalactic magnetic fields (IGMF). While we do not make any statements about these topics in this current work, as the catalog of high redshift TeV detections increases (with an accompanying increase in statistics on individual sources) our understanding of the properties of the EBL and IGMF will increase as well. A subsequent study of this source with these goals in mind will be the subject of a future publication.  Additionally, the next generation IACT array CTA \citep{Actis} (along with additional data collection with $\textit{Fermi}$-LAT to improve the MeV-GeV spectral measurement) will be able to much more precisely measure the high energy emission of distant objects such as 1ES 0414+009, providing valuable insights into the nature of some of the most distant TeV targets known as well as fundamental measurements of the EBL and IGMF. 

\section{Acknowledgements} 
This research is supported by grants from the U.S. Department of Energy Office of Science, the U.S. National Science Foundation and the Smithsonian Institution, by NSERC in Canada, by Science Foundation Ireland (SFI 10/RFP/AST2748) and by STFC in the U.K. We acknowledge the excellent work of the technical support staff at the Fred Lawrence Whipple Observatory and at the collaborating institutions in the construction and operation of the instrument. This work also made use of data supplied by the UK $\textit{Swift}$ Science Data Centre at the University of Leicester.

\bibliographystyle{apj}
\bibliography{refs}

\pagebreak

\begin{table}[center]
\centering 

\begin{center}
 \begin{tabular}{|c|c|c|c|c|}

 \hline
\textbf{Parameter}&  \textbf{SSC} & \textbf{EC}  & \textbf{Lepto-Hadronic} \\ \hline
$\gamma_{emin}$ &  2$\times$10$^{5}$  & 9$\times 10^{4}$    &   4.8$\times10^{3}$         	 \\\hline
$\gamma_{emax}$ &  5$\times10^{6}$  &  3$\times10^{6}$  &     1.0$\times10^{5}$       	 \\\hline
$q_{e} $ & 3.5   &  3.5  &    3.6        	 \\\hline
$\gamma_{pmin}$ &  - & -    &   1.0$\times10^{3}$         	 \\\hline
$\gamma_{pmax}$ &  - &  -  &     1.0$\times10^{11}$       	 \\\hline
$q_{p} $ & -   &  -  &        1.8    	 \\\hline
$\eta$ &120 &40.0  & 3\\\hline
B (G) &0.008 &0.044 & 30\\\hline
$\Gamma$ &40 &40 & 20\\\hline
R (cm) &2.1$\times10^{17}$ &7$\times10^{16}$&1.0$\times10^{16}$  \\\hline
$\Theta_{obs} (^{\circ})$ &1.43 &1.43 & 2.86 \\\hline
$\Delta t_{min. variability} (hrs)$ & 62.5 &20.9 &5.95  \\\hline
$T_{EC} (K)$ & - & 1000 & -  \\\hline
$\textit{u}_{EC} (erg/cm^{-3})$ & - & 3$\times$10$^{-9}$ & - \\\hline
$L_{e}$ (erg/s) &3.07$\times10^{44}$ &6.93$\times10^{43}$ &1.91$\times10^{41}$ \\\hline
$L_{B}$ (erg/s)&1.69$\times10^{43}$ &5.69$\times10^{43}$ & 1.35$\times10^{47}$\\\hline
$\epsilon$ &0.055 &0.82 & 7.06$\times10^{5}$ \\\hline
$L_{p}$ &- &- & 8.0$\times10^{43}$\\\hline
$\epsilon_{pB}$=L$_{B}$/L$_{p}$ &- & -& 1.6$\times10^{3}$\\\hline
$\epsilon_{pe}$=L$_{e}$/L$_{p}$ &- & -& 2.4$\times10^{-3}$\\\hline\end{tabular}

\caption{Summary of the emission zone parameters used for modeling the synchrotron self-Compton, external Compton, and lepto-hadronic models as described in the text.}
\end{center}
\end{table}

\begin{figure}[t]
\begin{center}
   \includegraphics[width=0.5\textwidth,height=0.5\textwidth]{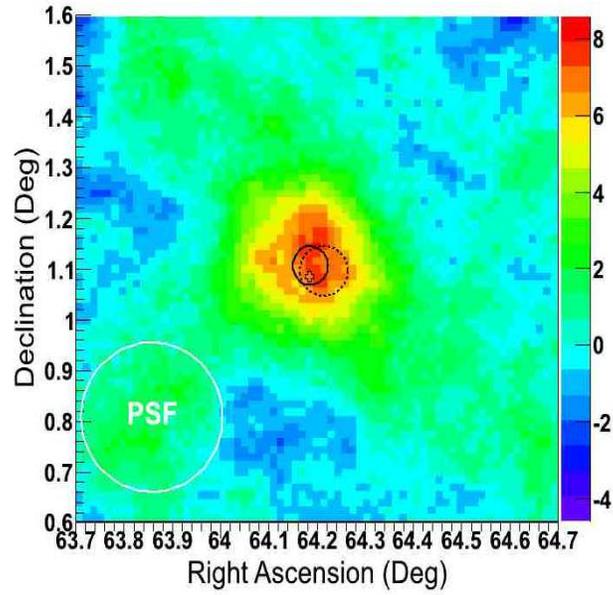}
\end{center}
\caption[]{The two-dimensional significance map of the 1ES 0414+009 region from VERITAS observations made at VHE gamma-ray energies with the color scale representing units of standard deviation of the corresponding excess. The cross represents the optical position of the AGN from \citet{0414OpticalPosition}, the solid circle represents the best fit VERITAS position with associated statistical and systematic errors, and the dotted circle represents the position of the excess observed by $\textit{Fermi}$-LAT (statistical error only). The white circle in the lower left corner represents the scale of the VERITAS point spread function of 0.12$^{\circ}$.}
\end{figure}

\begin{figure}[h]
\begin{center}
   \includegraphics[width=0.6\textwidth,height=0.5\textwidth]{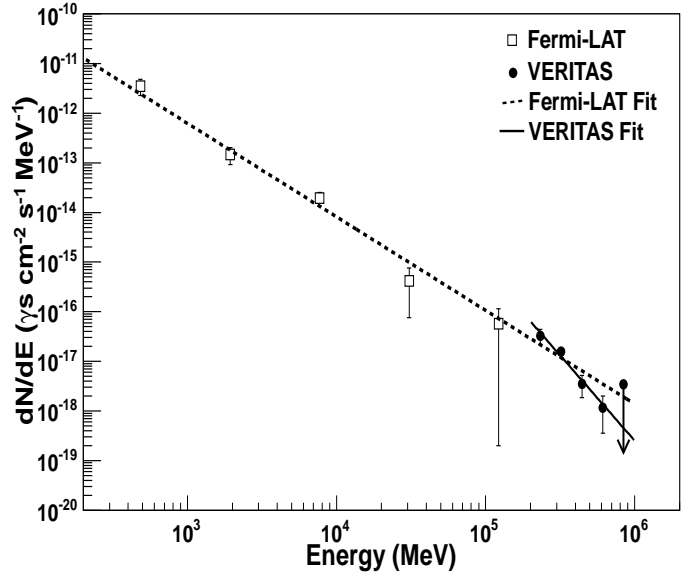}
\end{center}
\caption[]{The derived spectral points from both $\textit{Fermi}$-LAT and VERITAS observations (non-simultaneous) of 1ES 0414+009. The dotted line shows the power law fit to the $\textit{Fermi}$-LAT spectrum extrapolated to TeV energies, while the solid line shows the power-law fit to the observed VERITAS TeV spectrum.}
\end{figure}

\begin{figure}[h]
\begin{center}
   \includegraphics[width=\textwidth,height=0.45\textheight]{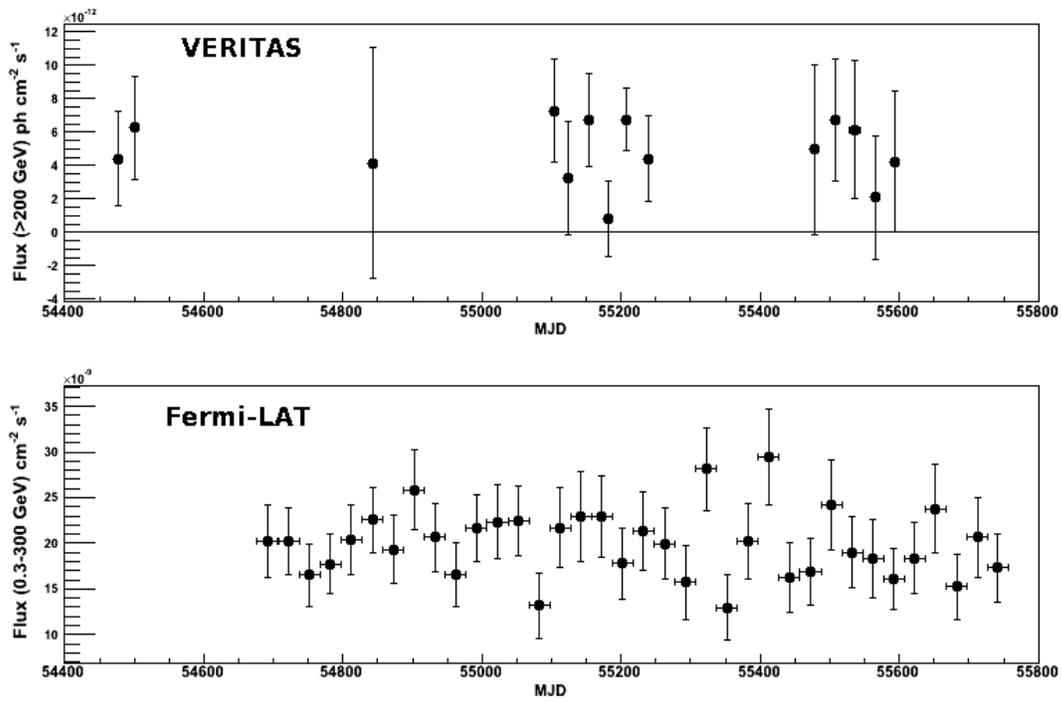}
\end{center}
\caption[]{The VERITAS (top) and \textit{Fermi}-LAT light curves for the observations detailed in this work. Errors shown on both light curves are statistical only.}
\end{figure}

\begin{figure}[h]
\begin{center}
   \includegraphics[width=\textwidth,height=0.65\textheight]{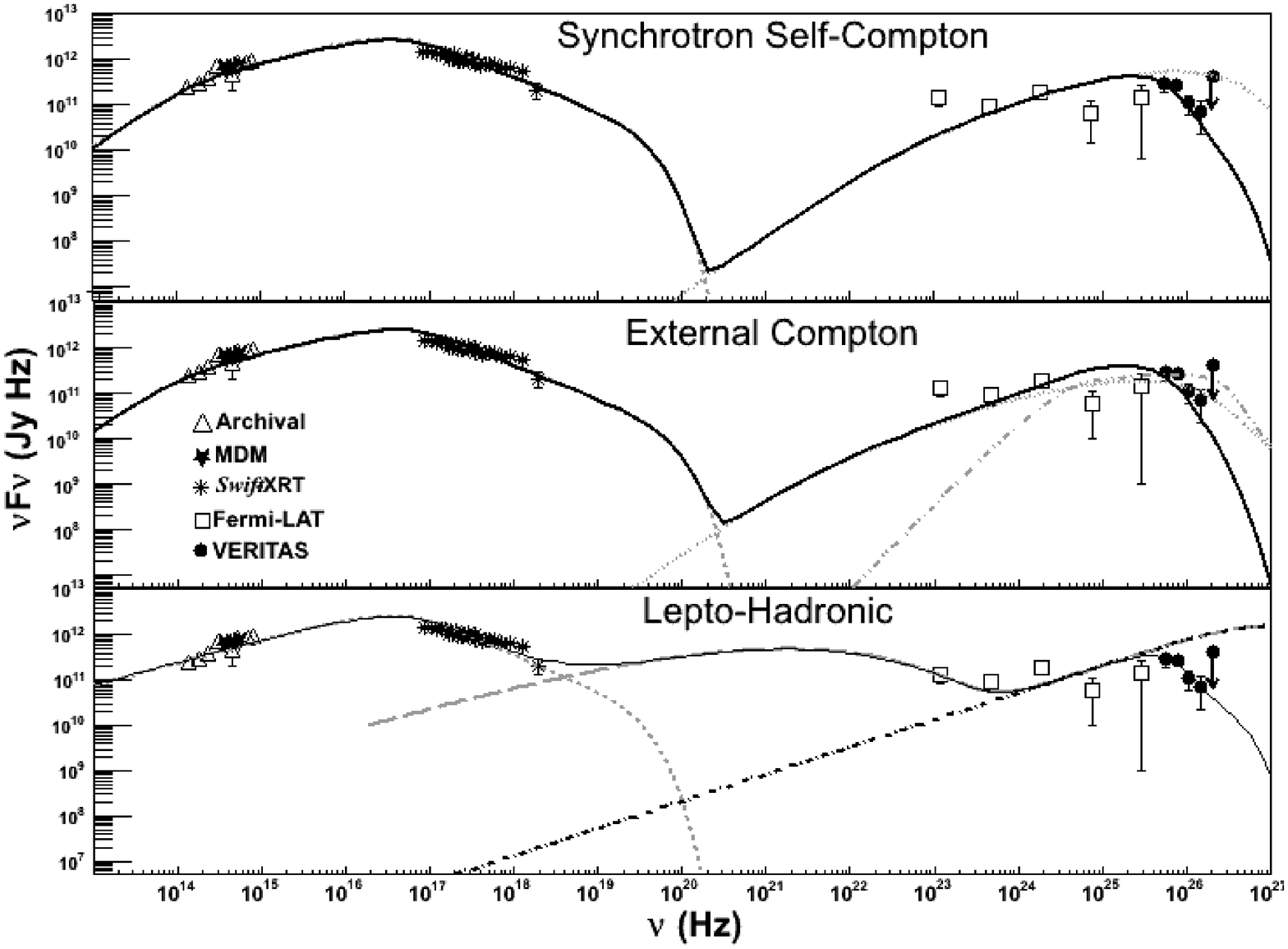}
\end{center}
\caption[]{The broadband SED constructed from the data sets described in the text. The top panel shows the SSC model results with contributions from primary synchrotron (dashed line), synchrotron self-Compton (dotted line), and EBL correct total (solid line). The middle panel shows the external Compton model results with contributions from primary synchrotron (dashed line), synchrotron self-Compton (dotted line), external Compton (dotted-dashed line), and EBL corrected total (solid line). The bottom panels shows the lepto-hadronic results with contributions from electron synchrotron (dotted line),  proton synchrotron (dotted-dashed), proton synchrotron with electromagnetic cascades (dashed line), and EBL corrected total.  }

\end{figure}

\end{document}